# Architecting Human-AI Cocreation for Technical Services – Interaction Modes and Contingency Factors

Jochen Wulf[1][0000-0001-5553-8850] , Jürg Meierhofer[1][0000-0003-1645-5686]
and Frank Hannich[1]

[1] ZHAW Zurich University of Applied Sciences, Gertrudstrasse 15, 8401 Winterthur, Switzerland

**Abstract.** Agentic AI systems, powered by Large Language Models (LLMs), offer transformative potential for value co-creation in technical services. However, persistent challenges like hallucinations and operational brittleness limit their autonomous use, creating a critical need for robust frameworks to guide human-AI collaboration. Drawing on established Human-AI teaming research and analogies from fields like autonomous driving, this paper develops a structured taxonomy of human-agent interaction.

Based on case study research within technical support platforms, we propose a six-mode taxonomy that organizes collaboration across a spectrum of AI autonomy. This spectrum is anchored by the Human-Out-of-the-Loop (HOOTL) model for full automation and the Human-Augmented Model (HAM) for passive AI assistance. Between these poles, the framework specifies four distinct intermediate structures. These include the Human-in-Command (HIC) model, where AI proposals require mandatory human approval, and the Human-in-the-Process (HITP) model for structured workflows with deterministic human tasks. The taxonomy further delineates the Human-in-the-Loop (HITL) model, which facilitates agent-initiated escalation upon uncertainty, and the Human-on-the-Loop (HOTL) model, which enables discretionary human oversight of an autonomous AI.

The primary contribution of this work is a comprehensive framework that connects this taxonomy to key contingency factors—such as task complexity, operational risk, and system reliability—and their corresponding conceptual architectures. By providing a systematic method for selecting and designing an appropriate level of human oversight, our framework offers practitioners a crucial tool to navigate the trade-offs between automation and control, thereby fostering the development of safer, more effective, and context-aware technical service systems.

**Keywords:** human in the loop, hybrid intelligence, human-autonomy teaming, human-agent interaction , technical service systems

## 1 Introduction

The adoption of agentic artificial intelligence (AI) is set to fundamentally reshape technical services, with recent industry projections indicating that AI will handle as much as 68% of customer service and support interactions by 2028 (Cisco, 2025). This transformation is driven by a technological evolution from simple chatbots to proactive AI agents, designed to act as a sophisticated "digital workforce" capable of executing complex workflows (Wolfe, 2025).

A recent survey by PagerDuty (2025) reveals that a vast majority of technology leaders are already experimenting with agentic AI, viewing it as a critical component for future operations. However, the same report highlights significant practitioner concerns, such as security and reliability calling for effective human oversight. This creates a critical design challenge. On one hand, the potential benefits are immense; studies show that augmenting human agents with AI can significantly increase productivity (Brynjolfsson et al., 2023). On the other hand, the known risks of generative AI, such as factual incorrectness, and the documented difficulties in achieving effective human-AI synergy mean that premature or poorly designed automation can degrade service quality and erode customer trust (Passerini et al., 2025; Vaccaro et al., 2024).

For managers and system designers, this landscape creates a clear and pressing need for a systematic framework to guide decisions on how to structure this collaboration. This leads to our central research question: How can different modes of human-agent collaboration be structured and selected to effectively support technical service delivery?

This paper addresses this question by employing a case study methodology to construct a comprehensive framework for designing human-agent systems. Informed by established theories on human-AI-cocreation, we analyze real-world use cases from leading technology providers to deduce a six-mode taxonomy of interaction. Our primary contribution is synthesizing these empirically derived modes with their underlying architectural principles and the key contingency factors that govern their use, such as task complexity and operational risk. The resulting framework provides practitioners with actionable design guidance to navigate the trade-offs between automation and human oversight, enabling the creation of more effective and reliable technical service systems.

## 2 Prior Research

The increasing integration of artificial intelligence into complex operational environments has catalyzed a shift in human-computer interaction research. The focus is moving beyond simple interfaces towards dynamic, collaborative ecosystems where humans and AI agents act as interdependent partners (Borghoff et al., 2025; Gil et al., 2020). This evolution has given rise to several related, yet distinct, concepts for structuring this collaboration, each with its own theoretical underpinnings and architectural implications.

### 2.1 Core Concepts in Human-AI-Cocreation

The most established concept in this domain is Human-in-the-Loop (HITL). In its common definition, a HITL system involves an AI agent that operates autonomously but is designed to recognize its own limitations, actively escalating to a human expert when faced with uncertainty or a task that exceeds its capabilities (Jakubik et al., 2025; Takerngsaksiri et al., 2025). The human is thus brought "in the loop" to handle exceptions, provide validation, or resolve ambiguity (Van Zoelen et al., 2023). This model is foundational for applications requiring high reliability, such as security screening (Huang et al., 2025) and moderating autonomous vehicle behavior (Gil et al., 2020; Kuru, 2022).

More recent research has broadened this perspective, leading to the paradigms of Hybrid Intelligence (HI) and Human-Autonomy Teaming (HAT). Hybrid Intelligence is defined as a synergistic system that aims to achieve complex goals by combining the complementary strengths of humans and AI, allowing them to co-evolve and learn from each other over time (Cao et al., 2024; Maletzki et al., 2024; Van Zoelen et al., 2023). This moves beyond the exception-handling nature of HITL to a more continuous and symbiotic partnership. Similarly, Human-Autonomy Teaming (HAT) explicitly frames the human-AI relationship as a team, drawing parallels with human-human teaming (Lyons et al., 2021). The key distinction lies in the conceptual shift from viewing AI as a tool (automation) to viewing it as a genuine teammate (autonomy), capable of agency, shared intent, and interdependent action (Lyons et al., 2021). Researchers at the Harvard Business School have recently shown that AI as a "cybernetic teammate" can not only improve productivity but also employee experience (Dell'Acqua et al., 2025).
Simmler and Frischknecht (2021) provide a valuable taxonomy to clarify these distinctions, proposing that human-machine collaboration should be analyzed along two independent axes: automation and technical autonomy. Automation describes the allocation of tasks, while technical autonomy describes the degree of independence the AI has in executing its assigned tasks. This dual-axis view explains why a system can be fully automated yet possess varying levels of autonomy, from a simple deterministic system to an open, adaptive one.

### 2.2 Prior Approaches to Architectures and Interaction Models

To realize these conceptual models, researchers have proposed various architectures and patterns. From a high-level, system-theoretical perspective, Borghoff et al.(2025) describe a layered architecture composed of "executive agents" for core computations, "observer agents" to bridge computations and presentation, and "surface agents" for user-facing interactions.

More pragmatically, van Zoelen et al. (2023) introduced the concept of Team Design Patterns (TDPs) to capture recurring modes of collaboration. Their work identified three primary patterns:
- AI Advisor and Human Performer: The AI analyzes options and provides recommendations, but the human makes the final decision.

- AI Performer and Human Assistant: The AI performs a task autonomously and requests human assistance only when it encounters a limitation. This pattern closely mirrors the classic HITL definition.
- AI Performer and Human Validator: The AI performs a task autonomously and provides an overview to a human supervisor, who verifies the information and provides feedback.

These patterns have been influential, with subsequent work aiming to operationalize them. For instance, Maletzki et al. (2024) illustrate how these TDPs can be realized using data-centric business process models, while Reinhard et al. (2024) instantiate the "AI Advisor" pattern as a supporter co-agent for frontline service employees. Application-specific frameworks also provide concrete architectural examples. Takerngsaksiri et al. (2025) present the HULA framework for software engineering, which employs an "AI Planner Agent" and an "AI Coding Agent" that collaborate with a human engineer in a multi-stage HITL process. In industrial robotics and autonomous driving, research has focused on control-theory models that enable AI to proactively cooperate with humans and mitigate limitations such as human reaction time-delays (Kuru, 2022; Qin et al., 2024). To manage the human workload in HITL systems, Jakubik et al. (2025) propose an "AI-in-the-Loop" (AIITL) model where "artificial experts" are created to learn from human experts, thereby reducing the need for repetitive manual review.

## 2.3 Human-AI Co-Creation in Technical Services

A significant stream of recent research focuses on applying collaborative models specifically to technical service domains, where the goal is not just task completion but the co-creation of value (Cao et al., 2024). This paradigm views the interaction as a synergetic process where the combined output of the human and AI is superior to what either could achieve alone.

In a large-scale study of generative AI deployed in a real-world customer support setting, Brynjolfsson et al. (2023) provide compelling evidence of this co-creative effect. They analyzed the staggered introduction of an AI-based conversational assistant for over 5,000 agents and found that access to AI increased worker productivity by 15% on average. Critically, these gains were driven by the AI's ability to augment human capabilities, not merely automate tasks. The AI system, which monitored customer chats and provided real-time suggestions, functioned by capturing and disseminating the tacit knowledge of the most effective agents.

This empirical evidence supports architectural concepts proposed elsewhere. For example, Reinhard et al. (2024) examine different designs for co-creation in frontline service, proposing a supporter co-agent (unidirectional advice) and a collaborator co-agent (interactive prompting). The mechanism for value co-creation is often a direct, reciprocal loop. Li et al. (2023) propose a novel HITL configuration for IT Service Management (ITSM) where support agents are incentivized to provide high-quality data labels (providing long-term value to the AI) in direct exchange for immediate utilitarian value in the form of better ticket suggestions (providing immediate value to the human).

This co-creative paradigm also extends to physical and industrial services. In Security Inspection Services, a different form of co-creation emerges, focused on enhancing safety. Huang et al. (2025) propose a hybrid system that combines the AI's strength in rapid, high-recall contraband screening with the human inspector's superior judgment in reducing false alarms, thereby co-creating a safer and more efficient outcome. In Smart Manufacturing, the focus is on designing collaboration that augments the operator's skills and enhances well-being, viewing the interaction as a holistic, socio-technical process (Hartikainen et al., 2024).

Across these diverse technical service domains, the unifying theme is a move away from simple task offloading towards designing symbiotic relationships. The goal is to leverage AI to reduce cognitive load, improve the quality and safety of service delivery, and enhance the capabilities and experience of the human operator through continuous, collaborative interaction (Brynjolfsson et al., 2023; Reinhard et al., 2024).

## 3 Method

### 3.1 Research Approach

This study employs a comparative case study methodology to investigate the structuring of human-agent collaboration in technical services. A case study approach is particularly well-suited for this research as it enables an in-depth, contextualized exploration of a contemporary phenomenon within its real-world setting (Yin, 2014). Given that agentic AI is a nascent field, this method allows for a rich understanding of the complex interplay between technology, processes, and human actors (Eisenhardt, 1989). By comparing multiple cases, we can identify recurring patterns and variations, thereby enhancing the external validity and generalizability of our findings (Eisenhardt, 1989; Yin, 2014).

### 3.2 Case Selection and Data Collection

The selection of cases was purposive, focusing on three leading providers of enterprise software platforms that offer advanced AI capabilities for technical service management. The selection was informed by an analysis of recognized market surveys that identify industry leaders (i.e., (Gartner, Inc., 2024) and (Gartner, Inc., 2025)). Choosing market-leading platforms ensures that our analysis is based on influential, state-of-the-art technologies that are representative of current and emerging industry practices. An overview of the selected cases is provided in Table 1.

**Table 1.** Cases Overview.

| Platform | Provider | Platform Description | Sources |
|---|---|---|---|
| Dynamics 365 Customer Service | Microsoft | An omnichannel customer service platform featuring an integrated AI "Copilot" to assist agents and automate resolutions. | (D365 Community, 2025; Microsoft, 2024, 2025; O'Quinn, 2024; Rand Group, 2025; Sahye, 2025; Talan, 2025) |
| Customer Service AI | Salesforce | A customer service and engagement platform utilizing "AgentForce" to generate responses, provide insights, and automate bot interactions. | (Goldman, 2024; Golubova, 2025; Salesforce, 2025a, 2025b, 2025c; Salesforce Help, 2025) |
| Now Platform | ServiceNow | An enterprise workflow automation platform with embedded generative AI ("Now Assist") for creating intelligent, self-service experiences and virtual agents. | (Roethof, 2025; ServiceNow, 2024, 2025c, 2025a; Winklix, 2025) |

Data for this study were collected between April and May 2025 from publicly available digital sources for each platform. These sources included official product webpages, technical documentation, promotional videos, white papers, and corporate blog posts that described and demonstrated the functionalities of their respective AI solutions for service contexts.

### 3.3 Data Analysis

We applied a theoretical coding lens derived from human-AI teaming research. We analyzed each identified activity to understand how responsibility and task allocation are organized between human and AI agents. The analysis was guided by two central questions:
- *Responsibility*: Who is the primary owner of this activity—the human, the AI, or is it a shared responsibility?
- *Task Allocation*: What is the trigger for allocating the task? Is it a predefined step in a workflow, an AI-initiated escalation due to uncertainty, or a discretionary action taken by a human supervisor?

Through an iterative process of coding and comparison across the four cases, distinct patterns of responsibility and task allocation emerged. These patterns were systematically grouped and conceptualized, forming the basis of the six-mode taxonomy of human-agent interaction presented in our results.

## 4 Results

Our cross-case analysis of technical service management delineates six distinct modes of human-AI collaboration. These modes are conceptualized along a spectrum of increasing system autonomy, mapping different configurations of human and AI involvement onto the standard process framework—which progresses from initial receipt and diagnosis, through solution formulation and approval, to final communication and case closure (Wulf & Winkler, 2020). This section presents our findings for each mode, beginning with those characterized by the most significant human involvement.

### 4.1 Human-Augmentation-Mode (HAM)

The Human-Augmented Model (HAM) represents the most basic level of human-AI collaboration, where the human expert retains complete control over the entire task. In this mode, the AI system functions as a passive, real-time assistant, providing a stream of potentially relevant information to augment the human agent's knowledge and awareness. As depicted in our process visualization (see Figure 1), all core workflow steps—from receiving the request to closing the case—are executed by the human agent. The AI operates as a supplementary entity, offering data and suggestions but never directly participating in the workflow or communicating with the customer.

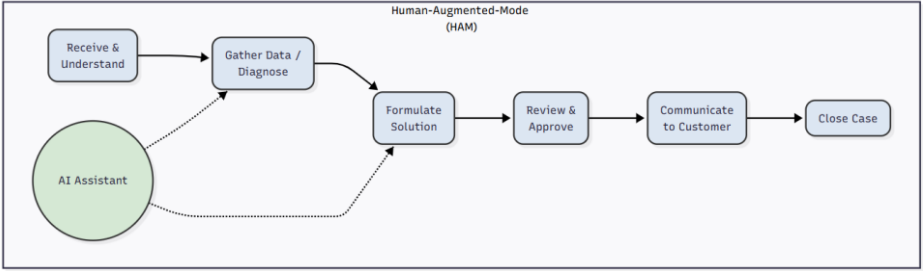

Fig 1: Human-Augmented-Mode (green=AI, blue=human)

A prime example of this model was identified in the functionalities of Microsoft Dynamics 365 Customer Service. In this use case, a human support engineer engages in a live conversation with a machine operator to diagnose a technical issue. The system's integrated Microsoft Dynamics 365 Copilot provides assistive capabilities designed to " Empower service representatives to resolve issues quickly" (Microsoft, 2025), Copilot's primary function is to provide real-time, contextual assistance. During a live chat, it offers agents dynamic summaries of the ongoing conversation, ensuring they have the essential context to provide informed support.
This augmentation is delivered through several specific features. First, the agent has access to an "Ask a question" tab, which allows the human agent to ask questions conversationally, and Copilot provides answers based on both internal knowledge bases and external trusted domains, effectively eliminating the need for manual information retrieval (Talan, 2025).

Second, the system provides intent-based and contextual suggestions. Instead of just matching keywords, the AI analyzes the live conversation to provide "contextual suggestions to agents on what they can do next based on the active case or customer conversation context" (Microsoft, 2024). This proactive guidance helps the engineer steer the diagnostic process by suggesting relevant questions to ask the customer, thereby reducing handling times.

Finally, Copilot can actively participate in response formulation. For instance, an agent can use a "Draft a Chat Response" feature, where Copilot drafts a reply based on the conversation's context. The human agent must then "review, edit if needed, and send the response" (D365 Community, 2025), placing the human in a clear command role over the AI-generated content.

This use case is a clear instantiation of the HAM framework. First, the human agent maintains complete command and is the sole communicator with the machine operator. The AI does not draft full responses or perform actions on the agent's behalf. Second, the AI's role is exclusively one of passive augmentation; it provides context, surfaces knowledge, and suggests questions, but the human agent is not obligated to use them. Finally, the model preserves the agent's autonomy and judgment. The engineer can leverage the AI's suggestions to guide their diagnosis or can choose to ignore them entirely based on their own expertise, perfectly aligning with the theoretical description of a human-led process enhanced by a passive AI assistant.

### 4.2 Human-in-Control (HIC)

The Human-in-Command (HIC) model represents a significant escalation in system autonomy compared to the HAM model. Here, the AI agent shifts from a passive informational role to an active propositional one. The system is capable of performing initial data gathering and formulating a complete, actionable solution or response. However, ultimate authority and accountability reside with the human operator, who must perform a mandatory review and provide explicit approval before any action is taken or communication is sent to the customer.

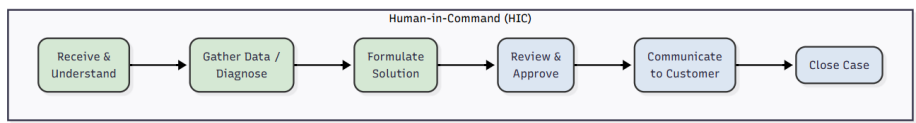

Fig 2: Human-In-Command (green=AI, blue=human)

As illustrated in our process framework (Figure 2), the HIC model assigns the initial workflow stages—Receiving & Understanding, Gathering Data, and Formulating the Solution—to the AI agent. This leverages the AI's speed in processing requests and drafting content. Crucially, the process flow then incorporates a mandatory handoff to the human agent for the critical stages of Review & Approve, Communicate to Customer, and Close Case. This ensures that no AI-generated output reaches the customer without explicit human validation.

A clear instantiation of the HIC model is observed in the capabilities of Salesforce's "Agentforce" product (Salesforce, 2025a). The workflow begins when an agent receives a customer query. The AI can produce summaries of the customer query and prior and surface relevant knowledge articles (Salesforce, 2025c), providing the agent with immediate context before any response is formulated.

Building on this understanding, the core HIC function is "Service Replies". The AI is designed to auto-generate personalised responses grounded in the knowledge base: "Receive AI-generated replies crafted from data from the conversation or from your company's trusted knowledge base. Enable service reps to share these replies with customers with one click or edit them before sending" (Salesforce, 2025b). Crucially, the workflow incorporates a mandatory human validation gate.

This architecture firmly places the human in command. The AI is leveraged for the speed and efficiency of drafting a complete, contextual response, but the system is explicitly designed so that a human agent performs the final, critical steps of verification and communication (Salesforce, 2025b).

## 4.3 Human-in-the-Process (HITP)

The Human-in-the-Process (HITP) model represents a paradigm of structured, deterministic collaboration. Unlike the preceding models, HITP is defined by a pre-planned division of labor within a single, integrated workflow. The process is designed to be predominantly autonomous, but it contains one or more mandatory steps that must be executed by a human. The trigger for human involvement is not an AI-initiated request for help (as in HITL) nor a proposal awaiting sign-off (as in HIC), but rather a predefined, non-negotiable step in the process map.

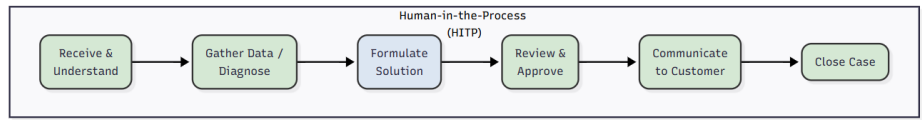

Fig 3: Human-in-the-Process (green=AI, blue=human)

Our process visualization (Figure 3) illustrates this concept by showing a workflow that is largely executed by the AI, but which is designed to halt and assign a specific operational task to a human agent. In this framework, the human is not a supervisor but an integrated operator, performing a function that the AI is not equipped or authorized to handle, such as tasks requiring high-stakes judgment or physical interaction.

This model is exemplified by the potential of enterprise workflow platforms like ServiceNow. In our composite use case, an AI-powered monitoring system perpetually oversees a piece of industrial machinery. Upon detecting a critical anomaly, the system autonomously performs the initial steps of the process: "When a monitoring system detects an anomaly or an issue […], ServiceNow can automatically create a high-priority incident, categorizing it based on business impact." (Winklix, 2025) At this point, the automated workflow is designed to pause. It then assigns a specific task to a human Service Manager: to review the situation and formally approve the dispatch of an on-site field technician. This approval step is deemed to require human-level strategic judgment concerning business impact, cost, and resource allocation. Once the manager provides the approval within the system, the workflow seamlessly resumes, with the AI proceeding to automatically handle the subsequent scheduling and logistics (ServiceNow, 2025a).

This example is a direct application of the HITP model because the human intervention is a required and pre-engineered part of the process flow. The platform's ability to create automated workflows that connect people, systems, and data (ServiceNow, 2025c) is the core enabler for this mode of collaboration. It allows organizations to combine the efficiency of end-to-end automation with the indispensable accountability of human judgment at critical, pre-identified junctures. The human's role is neither supervisory nor exceptional, but rather operational and integral to the successful execution of the automated process.

## 4.4 Human-in-the-Loop (HITL)

The Human-in-the-Loop (HITL) model describes a system of conditional autonomy, where an AI agent operates independently until it encounters a situation that exceeds its pre-defined operational capabilities or confidence thresholds. The defining characteristic of this model is that the AI system itself is responsible for recognizing its own limitations and actively initiating a handoff to a human expert. The human is thus brought "in the loop" on an exception basis to resolve ambiguity, handle complexity, or provide a final judgment.

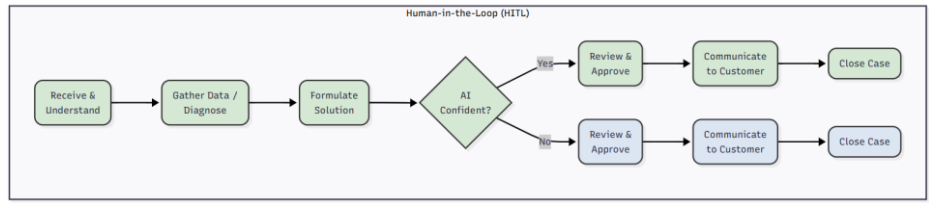

Fig 4: Human-in-the-Loop (green=AI, blue=human)

As illustrated in our process model (Figure 4), the HITL workflow includes a critical decision node. The AI proceeds autonomously through the initial steps of the process, but upon formulating a solution, it must pass through a confidence gate. If the system's confidence is high, it completes the remaining steps autonomously. However, if the confidence is low, the process flow is automatically rerouted, escalating the entire case to a human agent for resolution.

ServiceNow's Virtual Agent implements this HITL model by utilizing Natural Language Understanding (NLU) intent-confidence thresholds. When the confidence score for an interpreted intent falls below the configured threshold—typically around 60 percent—the system automatically initiates a fallback flow or escalates the conversation to a live agent (ServiceNow, 2025b). When escalation is triggered, the Virtual Agent executes script calls to transfer the session, provided that an agent is available (Roethof, 2025). The full conversation transcript and associated metadata are maintained in an interaction record, ensuring that the human agent receives complete context upon handoff.

For example, a technician may use the Virtual Agent to collect structured data such as error codes. When the technician describes sensory details—such as "a loud grinding noise coming from the main housing"—the NLU confidence is likely to fall below the threshold. In this event, the system automatically escalates the session to a human expert, preserving the transcript and context to ensure continuity and efficiency.

This ServiceNow use case adheres closely to HITL principles by operating autonomously until self-assessment detects low confidence, initiating escalation without manual intervention, and preserving full conversational context for human agents. The integration with live-agent infrastructure ensures that handoffs are timely and informationally rich.

## 4.5   Human-on-the-Loop (HOTL)

The Human-on-the-Loop (HOTL) model introduces a paradigm of supervisory control over a fully autonomous agent. In contrast to models requiring direct human participation, the AI in an HOTL system is designed to execute the entire workflow from start to finish without planned intervention. The human's role is transformed from an operator to a supervisor, who monitors the AI's performance—often through high-level indicators—and can proactively intervene at their own discretion. The trigger for intervention is not a system request but the human's own judgment.

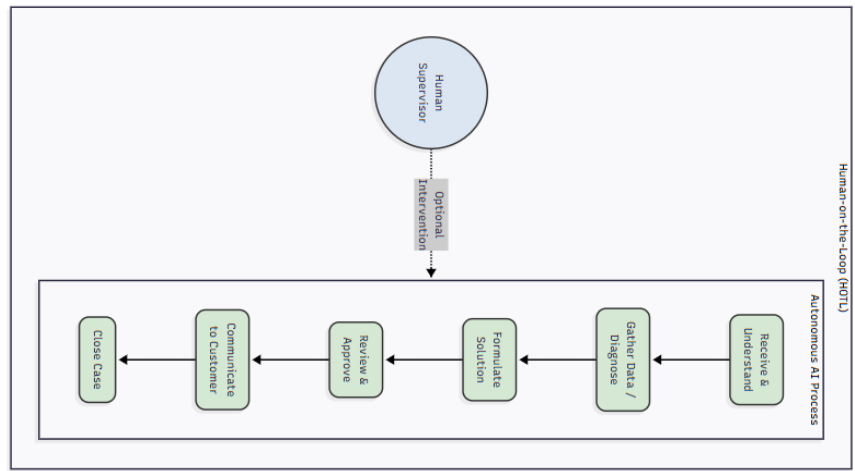

Fig 5: Human-on-the-Loop (green=AI, blue=human)

Our conceptual process model (Figure 5) visualizes this relationship distinctly. The core workflow is depicted as a self-contained, end-to-end process executed entirely by the AI. The human agent is positioned externally to this flow, representing their supervisory status. A dotted line indicates their capability to optionally intervene in the process at any stage, an action that is based on their initiative rather than a system requirement.

Salesforce Service Cloud clearly embodies HOTL through its Live AI Agent Monitoring (Salesforce Help, 2025). Supervisors gain real-time visibility into conversations managed by Einstein Service Agents—now surfaced in the new "AI Agents" tab of Omni-Supervisor—where they can view live transcripts, track sentiment flags, and proactively transfer chats to a human agent when their judgment deems it necessary (Golubova, 2025).

This approach aligns with Salesforce's commitment to putting "humans at the helm" of its AI technologies. As Salesforce puts it, "the AI revolution is not just about technological innovation — it's also about empowering humans to sit successfully at the helm of AI" (Goldman, 2024). Their Agentforce platform further strengthens this supervisory model, offering a Command Center that gives leaders full observability and real-time control over autonomous AI agents (Golubova, 2025; Salesforce Help, 2025). Together, these features deliver a true HOTL experience: AI handles workflows end-to-end, while humans supervise strategically and choose when to step in—without any automatic escalation mechanism, only human-initiated intervention.

### 4.6 Human-out-of-the-Loop (HOOTL)

The Human-Out-of-the-Loop (HOOTL) model represents the extreme of the autonomy spectrum. In this mode, the AI system is entrusted with executing the entire operational process from initiation to completion without any requirement for human intervention, supervision, or approval. The system not only handles the execution of tasks but often also the initial detection of the need for action, embodying a shift from reactive service to proactive or predictive operations.

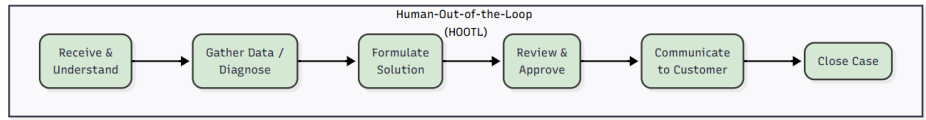

Fig 6: Human-out-of-the-Loop (green=AI, blue=human)

As illustrated in our process framework (Figure 6), the HOOTL model is represented by an unbroken, linear sequence of tasks performed entirely by the AI. There are no human agents integrated into the workflow, nor are there external supervisors. The process is self-contained and self-sufficient, requiring a very high degree of system reliability, a predictable operational domain, and a low tolerance for the types of errors that would necessitate human oversight.

Microsoft Dynamics 365 Field Service exemplifies the HOOTL model in predictive maintenance scenarios (O'Quinn, 2024). In this model, the AI system autonomously manages the entire service process—from detecting potential issues to dispatching technicians—without human intervention. This approach is particularly evident in scenarios where equipment failures are anticipated through predictive analytics. For instance, when IoT sensors integrated with Dynamics 365 detect anomalies such as increased vibration in machinery, the system's AI capabilities, including Copilot, analyze this data to predict impending failures (O'Quinn, 2024). Upon identifying a potential issue, the AI autonomously generates a work order, schedules a technician with the appropriate skills and proximity, and orders necessary parts through integrated supply chain management (Rand Group, 2025; Sahye, 2025). Throughout this process, the system operates without human oversight, embodying the HOOTL model.

In summary, Microsoft Dynamics 365 Field Service demonstrates a practical application of the HOOTL model by leveraging AI and IoT technologies to autonomously manage predictive maintenance workflows, thereby reducing the need for human intervention and enhancing operational efficiency.

## 5 Discussion

### 5.1 Contingency Factors for Selecting a Collaboration Mode

Our analysis reveals that the selection of an appropriate human-agent interaction mode is not an arbitrary choice but a strategic decision contingent upon a set of interconnected factors. The optimal level of autonomy depends on the specific context of the task, the capabilities of the AI system, and the intended role of the human operator. Drawing from our cross-case analysis and the synthesized framework, we identify several key contingency factors that guide this selection.

*Task Complexity and Novelty:* The inherent nature of the task is a primary determinant. At one extreme, the HOOTL model is best suited for low-complexity, highly repetitive, and well-defined tasks where the range of variation is minimal. The Microsoft Dynamics 365 Field Service example (O'Quinn, 2024; Rand Group, 2025), which automates the predictable process of ordering a part based on sensor data, is a perfect illustration. At the other extreme, the HIC and HAM models are reserved for high-complexity and novel tasks. The HIC model, exemplified by Salesforce's draft review (Salesforce, 2025b, 2025c), is applied when the AI can attempt a solution but human judgment is essential to ensure correctness and nuance. The HAM model is used when the task is so novel or complex that the human must lead entirely, using the AI only for informational support. Intermediate models like HITL and HOTL are suited for moderate complexity; for instance, the ServiceNow HITL case (Roethof, 2025; ServiceNow, 2025b) is triggered when a novel query (a "grinding noise") outside the AI's known variations occurs.

*Safety, Criticality, and Risk:* The potential consequences of an error are a critical consideration. The HOOTL model is only viable for very low-risk, non-critical tasks where an error has minimal and easily reversible consequences. Especially in B2B environments, safety and risk have a direct impact on value creation and can lead to significant value destruction if not properly managed. The highest-risk and safety-critical tasks demand the HIC model. Providing technical instructions for machine operation, as in the Salesforce example (Salesforce, 2025b, 2025c), carries inherent risk, mandating that every output is human-validated before being sent to an operator. The intermediate models serve as a safety net. The HITL model (ServiceNow) (Roethof, 2025; ServiceNow, 2025b) mitigates risk by escalating unknown problems, while the HOTL model allows a human supervisor to prevent reputational or operational risk by intervening in a failing autonomous interaction.

*System Reliability and Trust:* The perceived reliability of and trust in the AI system directly correlates with the degree of autonomy granted. True HOOTL automation requires exceptionally high system reliability (>95% accuracy) and operator trust in its capabilities. Conversely, the HIC model is chosen when there is low trust in the AI's autonomous reasoning and outputs are treated as a "black box" needing verification. Trust in the HITP model is nuanced: trust is high for the automated steps of the workflow (e.g., ServiceNow creating a ticket (ServiceNow, 2025c; Winklix, 2025)) but is intentionally zero for the specific step that requires a human expert's judgment. The confidence scores in HITL and HOTL systems serve as a mechanism to build moderate trust by providing some transparency into the AI's internal state.

*Human Operator State (Workload, Fatigue, and Vigilance):* Finally, the impact on the human operator must be considered. The primary goal of the HOOTL model is to entirely eliminate the human workload for a specific task. Conversely, the HIC model imposes the highest cognitive workload and vigilance demand per item, as every AI output must be carefully scrutinized. This is best suited for lower-volume queues where dedicated human attention is feasible. The HOTL model requires moderate but continuous vigilance, creating a risk of automation bias if the system is overwhelmingly reliable. The HITL model offers a different trade-off: it demands very low vigilance until an escalation occurs, at which point it requires a short burst of high-intensity cognitive work from the expert. This makes it efficient for managing a high volume of standard requests punctuated by rare, complex exceptions.

## 5.2 Contributions to Research and Practice

This paper makes three targeted theoretical contributions by synthesizing prior research and addressing specific, documented gaps in the literature. First, we address the lack of standardized methods for operationalizing Team Design Patterns (Van Zoelen et al., 2023), which has led to bespoke, domain-specific architectures (Takerngsaksiri et al., 2025). We contribute a six-mode taxonomy (HAM, HIC, HITP, HITL, HOTL, HOOTL) that provides a reusable, technology-agnostic framework for translating these patterns into concrete architectural primitives.

Second, we provide a unified framework that connects architectural choices to specific operational contexts, a linkage that is currently absent. We achieve this by systematically mapping our six modes to key contingency factors such as task complexity and risk, providing practitioners with the clear, context-sensitive design guidance needed for implementation.

Finally, in response to the challenge of evaluating complex systems and achieving "scalable oversight" (Bowman et al., 2022), our framework serves as a foundational analytical tool. It enables a systematic comparison of different human-AI configurations and their trade-offs, laying the groundwork for more robust and standardized evaluation methods.

Furthermore, this research offers significant practical contributions for technology leaders, system designers, and managers tasked with integrating agentic AI into technical service operations. First, the six-mode taxonomy provides a much-needed common vocabulary for structuring design conversations. Instead of ambiguous discussions about AI's role, teams can use this classification to precisely define the level of human oversight and control for any given process. This clarity is invaluable for strategic planning, requirements definition, and communicating with stakeholders about how a new system will operate. Second, the framework provides a pragmatic decision-making tool by linking each collaboration mode to clear contingency factors like task complexity and operational risk. This directly addresses the concerns about reliability and the need for human oversight highlighted by technology leaders (PagerDuty, 2025). Managers can use this framework to conduct a structured risk assessment, justifying the choice to, for example, implement a restrictive Human-in-Command model for critical tasks versus a more efficient Human-on-the-Loop model for routine ones. Ultimately, this framework equips practitioners to move beyond ad-hoc experimentation. It provides a systematic approach to navigate the central trade-off between harnessing the productivity gains of AI (Brynjolfsson et al., 2023) and mitigating the inherent risks of premature or poorly designed automation (Vaccaro et al., 2024), leading to more reliable and effective technical service systems.

# 6  Conclusion

This paper introduces a six-mode taxonomy for human-agent collaboration in technical services, providing a framework that connects different interaction models to key contingency factors like task complexity and operational risk. This systematic approach offers a valuable tool for designing and implementing effective, context-aware human-AI systems.

The conclusions of this study are based on a qualitative analysis of publicly available data from four major technology providers. Future research could enhance the generalizability of these findings by including a broader and more diverse range of companies and AI platforms. Additionally, the rapid evolution of AI means that the presented taxonomy may need to be updated as new capabilities and interaction modes emerge.

Several avenues for future research are apparent. First, empirical studies are needed to quantitatively evaluate the performance and usability of each of the six proposed modes in real-world operational environments. Such research could measure key metrics like task completion time, error rates, and human operator satisfaction to validate and refine the presented framework.

Second, the sociopsychological implications of human-AI teaming merit deeper investigation. A reciprocal relationship likely exists between organizational culture and the selected collaboration mode; culture can drive the adoption of a particular model, which in turn may reshape workplace norms and values. Consequently, understanding the acceptance of these modes among both employees and customers is a critical area for research. Future studies should explore how different interaction models affect key human factors such as trust in the AI, the cognitive load imposed on agents, and the evolution of their skill sets. Comprehending the long-term effects of these systems on the workforce is therefore crucial for achieving a responsible and sustainable integration of AI into technical services.

Third, future research should aim to develop frameworks for selecting the optimal human-AI interaction mode through a risk-benefit analysis, one that weighs the potential for value creation against specific contextual factors of safety and risk.

Finally, there is an opportunity to develop and test "symbiotic" AI systems that can dynamically adapt their level of autonomy based on the evolving context of a task and the real-time state of the human user. This would represent a significant step beyond the current, more static models of collaboration.

# References

Borghoff, U. M., Bottoni, P., & Pareschi, R. (2025). A System-Theoretical Multi-agent Approach to Human-Computer Interaction. In A. Quesada-Arencibia, M. Affenzeller, & R. Moreno-Díaz (Eds.), *Computer Aided Systems Theory – EUROCAST 2024* (pp. 23–32). Springer Nature Switzerland. https://doi.org/10.1007/978-3-031-82949-9_3

Bowman, S. R., Hyun, J., Perez, E., Chen, E., Pettit, C., Heiner, S., Lukošiūtė, K., Askell, A., Jones, A., Chen, A., Goldie, A., Mirhoseini, A., McKinnon, C., Olah, C., Amodei, D., Amodei, D., Drain, D., Li, D., Tran-Johnson, E., … Kaplan, J. (2022). *Measuring Progress on Scalable Oversight for Large Language Models* (No. arXiv:2211.03540). arXiv. https://doi.org/10.48550/arXiv.2211.03540

Brynjolfsson, E., Li, D., & Raymond, L. R. (2023). *Generative AI at work*. National Bureau of Economic Research.

Cao, L., Lindman, J., Cronholm, S., & Göbel, H. (2024). *Design Principles for Hybrid Decision Support Systems in IT Service Management*. https://aisel.aisnet.org/ecis2024/track23_designresearch/track23_designresearch/11/

Cisco. (2025, May). *Agentic AI Poised to Handle 68% of Customer Service and Support Interactions by 2028*. https://newsroom.cisco.com/c/r/newsroom/en/us/a/y2025/m05/agentic-ai-poised-to-handle-68-of-customer-service-and-support-interactions-by-2028.html

D365 Community. (2025, July). *Effortless Chat Replies with Dynamics 365 Copilot*. https://www.powercommunity.com/effortless-chat-replies-with-dynamics-365-copilot/

Dell'Acqua, F., Ayoubi, C., Lifshitz-Assaf, H., Sadun, R., Mollick, E. R., Mollick, L., Han, Y., Goldman, J., Nair, H., Taub, S., & Lakhani, K. R. (2025). *The Cybernetic Teammate: A Field Experiment on Generative AI Reshaping Teamwork and Expertise* (SSRN Scholarly Paper No. 5188231). Social Science Research Network. https://doi.org/10.2139/ssrn.5188231

Eisenhardt, K. M. (1989). Building Theories from Case Study Research. *The Academy of Management Review*, *14*(4), 532. https://doi.org/10.2307/258557

Gartner, Inc. (2024). *Magic Quadrant for CRM Customer Engagement Center* (Technical Report No. 6006703). Gartner, Inc. https://www.gartner.com/en/documents/6006703

Gartner, Inc. (2025). *Market Guide for Field Service Management* (No. 6311147). Gartner, Inc. https://www.gartner.com/en/documents/6311147

Gil, M., Albert, M., Fons, J., & Pelechano, V. (2020). Engineering human-in-the-loop interactions in cyber-physical systems. *Information and Software Technology*, *126*, 106349.

Goldman, P. (2024, March). *Trusted AI Needs a Human at the Helm*. https://www.salesforce.com/news/stories/human-at-the-helm/

Golubova, S. (2025, February). *Announcing Live Monitoring of AI Agent for Supervisors*. https://unofficialsf.com/announcing-live-monitoring-of-ai-agent-for-supervisors/

Hartikainen, M., Spurava, G., Väänä, & Nen, K. (2024). Human-AI Collaboration in Smart Manufacturing: Key Concepts and Framework for Design. In *HHAI 2024: Hybrid Human AI Systems for the Social Good* (pp. 162–172). IOS Press. https://doi.org/10.3233/FAIA240192

Huang, Y., Wang, X., Zhang, Y., Chen, L., & Zhang, H. (2025). Application of human-in-the-loop hybrid augmented intelligence approach in security inspection system. *Frontiers in Artificial Intelligence*, *8*, 1518850.

Jakubik, J., Weber, D., Hemmer, P., Vössing, M., & Satzger, G. (2025). Improving the Efficiency of Human-in-the-Loop Systems: Adding Artificial to Human Experts. In D. Beverungen, C. Lehrer, & M. Trier (Eds.), *Solutions and Technologies for Responsible Digitalization* (pp. 131–147). Springer Nature Switzerland. https://doi.org/10.1007/978-3-031-80122-8_9

Kuru, K. (2022). *Technical report: Analysis of intervention modes in human-in-the-loop (hitl) teleoperation with autonomous ground vehicle systems*. https://clok.uclan.ac.uk/id/eprint/49575/

Li, M. M., Löfflad, D., Reh, C., & Oeste-Reiß, S. (2023). Towards the design of hybrid intelligence frontline service technologies–A novel human-in-the-loop configuration for human machine interactions. *Hawaii International Conference on System Sciences (HICSS)*, 332–341. https://www.alexandria.unisg.ch/bitstreams/dc2c54d3-04f5-4b77-bf1f-9677acb2616e/download

Lyons, J. B., Sycara, K., Lewis, M., & Capiola, A. (2021). Human–autonomy teaming: Definitions, debates, and directions. *Frontiers in Psychology*, *12*, 589585.

Maletzki, C., Rietzke, E., & Bergmann, R. (2024). Empowering large language models in hybrid intelligence systems through data-centric process models. *Proceedings of the AAAI Symposium Series*, *3*(1), 167–174. https://ojs.aaai.org/index.php/AAAI-SS/article/view/31196

Microsoft. (2024, October 10). *Get automatic prompts from Copilot*. https://learn.microsoft.com/en-us/dynamics365/release-plan/2024wave1/service/dynamics365-customer-service/get-automatic-prompts-copilot

Microsoft. (2025). *Dynamics 365 Customer Service*. https://www.microsoft.com/en-us/dynamics-365/products/customer-service

O'Quinn, S. (2024). *Harnessing transformative AI for field service excellence*. https://www.microsoft.com/en-us/dynamics-365/blog/business-leader/2024/09/26/harnessing-transformative-ai-for-field-service-excellence/

PagerDuty. (2025). *Agentic AI Survey 2025: Balancing Innovation and Readiness*. PagerDuty. https://www.pagerduty.com/wp-content/uploads/2025/03/Agentic-AI-Survey-Report_FINAL.pdf

Passerini, A., Gema, A., Minervini, P., Sayin, B., & Tentori, K. (2025). Fostering effective hybrid human-LLM reasoning and decision making. *Frontiers in Artificial Intelligence*, 7, 1464690.

Qin, Z., Wu, H.-N., & Wang, J.-L. (2024). Proactive cooperative consensus control for a class of human-in-the-loop multi-agent systems with human time-delays. *Neurocomputing*, 581, 127485.

Rand Group. (2025). *Microsoft Dynamics 365 Field Service Features and Capabilities*. https://www.rand-group.com/insights/microsoft/dynamics-365-field-service-features-and-capabilities/

Reinhard, P., Neis, N., Kolb, L., Wischer, D., Li, M. M., Winkelmann, A., Teuteberg, F., Lechner, U., & Leimeister, J. M. (2024). Augmenting Frontline Service Employee Onboarding via Hybrid Intelligence: Examining the Effects of Different Degrees of Human-GenAI Interaction. In M. Mandviwalla, M. Söllner, & T. Tuunanen (Eds.), *Design Science Research for a Resilient Future* (pp. 384–397). Springer Nature Switzerland. https://doi.org/10.1007/978-3-031-61175-9_26

Roethof, M. (2025). *50\+ (Un)documented Virtual Agent variables (vaInputs, vaVars, vaContext, vaSystem)*. https://www.servicenow.com/community/virtual-agent-nlu-articles/50-un-documented-virtual-agent-variables-vainputs-vavars/ta-p/2310088

Sahye, K. (2025). *AI in Dynamics 365 Field Service*. https://www.gestisoft.com/en/blog/ai-in-dynamics-365-field-service

Salesforce. (2025a). *Agentforce: The AI Agent Platform*. https://www.salesforce.com/agentforce/

Salesforce. (2025b). *Service AI*. https://www.salesforce.com/eu/service/ai/

Salesforce. (2025c, March). *Agentforce resolves 85 % of visitor issues on the Salesforce Help site*. https://www.salesforce.com/customer-stories/agentforce-for-customer-support/

Salesforce Help. (2025). *Monitor Real-Time Conversations Between AI Agent and Customers*. https://help.salesforce.com/s/articleView?id=release-notes.rn_asa_monitor.htm&release=254&type=5

ServiceNow. (2024). *Support Agent Productivity with GenAI*. https://www.servicenow.com/blogs/2024/support-agent-productivity-genai

ServiceNow. (2025a). *Field Service Management*. https://www.servicenow.com/products/field-service-management.html

ServiceNow. (2025b). *NLU Model Settings*. https://www.servicenow.com/docs/bundle/xanadu-intelligent-experiences/page/administer/natural-language-understanding/concept/nlu-model-settings.html

ServiceNow. (2025c). *Workflow Automation on the Now Platform*. https://www.servicenow.com/now-platform/workflow-automation.html

Simmler, M., & Frischknecht, R. (2021). A taxonomy of human–machine collaboration: Capturing automation and technical autonomy. *Ai & Society*, 36(1), 239–250.

Takerngsaksiri, W., Pasuksmit, J., Thongtanunam, P., Tantithamthavorn, C., Zhang, R., Jiang, F., Li, J., Cook, E., Chen, K., & Wu, M. (2025). *Human-In-the-Loop Software Development Agents* (No. arXiv:2411.12924). arXiv. https://doi.org/10.48550/arXiv.2411.12924

Talan. (2025, May). *Using Copilot in Microsoft Dynamics 365 Customer Service*. https://www.talan.com/americas/en/using-copilot-microsoft-dynamics-365-customer-service

Vaccaro, M., Almaatouq, A., & Malone, T. (2024). When combinations of humans and AI are useful: A systematic review and meta-analysis. *Nature Human Behaviour*, 8(12), 2293–2303.

Van Zoelen, E., Mioch, T., Tajaddini, M., Fleiner, C., Tsaneva, S., Camin, P., Gouvêa, T. S., Baraka, K., De Boer, M. H., & Neerincx, M. A. (2023). Developing team design patterns for hybrid intelligence systems. In *HHAI 2023: Augmenting Human Intellect* (pp. 3–16). IOS Press. https://ebooks.iospress.nl/volumearticle/63319?utm_source=hbo-kennisbank.nl&utm_content=link

Winklix. (2025). *Optimizing Incident Management with ServiceNow: Best Practices and Automation*. https://www.winklix.com/blog/optimizing-incident-management-with-servicenow-best-practices-and-automation/

Wolfe, B. (2025, March). *From copilots to agents: The evolution of generative AI is creating a new digital workforce*. https://www.businessinsider.com/generative-ai-evolution-software-companies-develop-ai-agents-workforce-2025-3

Wulf, J., & Winkler, T. J. (2020). Evolutional and transformational configuration strategies: A rasch analysis of IT providers' service management capability. *Journal of the Association for Information Systems*, *21*(3), 574–606.

Yin, R. K. (2014). *Case study research: Design and methods (applied social research methods)*.